# Jet – ISM Interaction in NGC 1167/B2 0258+35, a LINER with an AGN Past

G. Fabbiano[1], A. Paggi[2,3,4], R. Morganti[5,6], M. Baloković[7,8], M. Elvis[1], D. Mukherjee[9], M. Meenakshi[9], A. Siemiginowska[1], S. M. Murthy[10], T.A. Oosterloo[5,6], A. Y. Wagner[11], G. Bicknell[12]

1. Center for Astrophysics | Harvard & Smithsonian, 60 Garden St., Cambridge MA 02138, USA
2. Dipartimento di Fisica, Università degli Studi di Torino, via Pietro Giuria 1, I-10125 Torino, Italy
3. INAF-Osservatorio Astrofisico di Torino, via Osservatorio 20, I-10025 Pino Torinese, Italy
4. Istituto Nazionale di Fisica Nucleare, Sezione di Torino, via Pietro Giuria 1, I-10125 Torino, Italy
5. ASTRON, Netherlands Institute for Radio Astronomy, Oude Hoogeveensedijk 4, 7991 PD Dwingeloo, The Netherlands
6. Kapteyn Astronomical Institute, Groningen University, Postbus 800, 9700 AV Groningen, The Netherlands
7. Yale Center for Astronomy & Astrophysics, 52 Hillhouse Avenue, New Haven, CT 06511, USA
8. Department of Physics, Yale University, P.O. Box 208120, New Haven, CT 06520, USA
9. Inter-University Centre for Astronomy and Astrophysics, Pune 411007, India
10. JIVE: Joint Institute for VLBI ERIC, Oude Hoogeveensedijk 4, 7991 PD Dwingeloo, The Netherland
11. University of Tsukuba, Center for Computational Sciences, 1-1-1 Tennodai, Tsukuba, Ibaraki, 305-8577 Japan
12. Research School of Astronomy and Astrophysics, The Australian National University, Canberra, ACT 2611, A

Abstract

We report the results of joint Chandra/ACIS - NuSTAR deep observations of NGC 1167, the host galaxy of the young radio jet B2 0258+35. In the ACIS data we detect X-ray emission, extended both along and orthogonal to the jet. At the end of the SE radio jet, we find lower-energy X-ray emission that coincides with a region of CO turbulence and fast outflow motions. This suggests that the hot Interstellar Medium (ISM) may be compressed by the jet and molecular outflow, resulting in more efficient cooling. Hydrodynamic simulations of jet-ISM interaction tailored to NGC 1167 are in agreement with this conclusion and with the overall morphology and spectra of the X-ray emission. The faint hard nuclear source detected with Chandra and the stringent NuSTAR upper limits on the harder X-ray emission show that the active galactic nucleus (AGN) in NGC 1167 is in a very low-accretion state. However, the characteristics of the extended X-ray emission are more consonant to those of luminous Compton Thick AGNs, suggesting that we may be observing the remnants of a past high accretion rate episode, with sustained strong activity lasting ~ 2 x 10³ yr. We conclude that NGC1167 is presently a LINER, but was an AGN in the past, given the properties of the extended X-ray emission and their similarity with those of CT AGN extended emission.

# 1. Introduction

Super massive black holes (SMBH) are invoked by cosmological simulations to prevent the piling up of cold gas in the center of galaxies, thus regulating star formation and the growth of the SMBH itself (AGN feedback). Radio jets could be an efficient vector of this interaction, especially in their initial expansion (e.g., Begelman et al. 1984; Bicknell et al. 1997). These predictions, however, are difficult to test observationally in detail, because young radio jets tend to be compact, requiring high angular resolution observations. Here we report the results of our deep (200 ks) Chandra ACIS-S observations of the young radio jet B2 0258+35 hosted in the galaxy NGC 1167, which we obtained to investigate these issues. Of the sample of known young radio galaxies observed in X-rays (Siemiginowska et al. 2016), NGC 1167 is the only one that can be spatially resolved with Chandra to give us an X-ray view of the interaction within the central ~300 pc.

The host galaxy, NGC 1167, is an early-type (SA0) galaxy (z=0.0165, D~70 Mpc; 1''~300 pc), with a stellar radius of 16.3 kpc (Pandya et al. 2017). NGC 1167 is particularly rich in gas, including a large HI disk (~160 kpc diameter, Struve et al. 2010), molecular gas (O'Sullivan et al. 2015, Murthy et al. 2022) and warm ionized gas (Gomes et al. 2016). NGC 1167 is part of the low luminosity AGN sample of Ho et al. (2009) and is classified as a LINER in the 2" wide (600 pc) optical spectral slit. More recent IFU observations confirm the LINER nature of the circumnuclear emission and exclude that it may originate from stellar ionization, requiring widespread shocks in the inner 2" (Gomes et al. 2016). The mass of the SMBH in NGC 1167 is estimated to be 4.4 × $10^8$ $M_\odot$ (Kormendy & Ho 2013).

B2 0258+35 is a low luminosity, ~ 0.4 Myr old, Compact Steep Spectrum (CSS) radio source (Giroletti et al. 2005; Shulevski et al. 2012; Brienza et al 2018), and there is evidence from optical and molecular studies that the jet is interacting with the galaxy ISM (Murthy et al. 2022 and refs therein). Giroletti et al. (2005) report that the inner ~3" diameter CSS source has a radio luminosity L(408 MHz)=2.3 x $10^{24}$ WHz$^{-1}$, at the higher end of typical radio quiet AGNs; they also find that the radio source is ~1 kpc in projected size, consisting of a core and two jets. The northern jet is comparatively faint, while the southern jet is bent sharply about 0.5 kpc from the core. This asymmetry both in brightness and morphology suggests an interaction with the ISM. At larger radii, well outside the optical extent of the galaxy, is an older, age ~100 Myr, double-lobed radio source extending to ~13' (~240 kpc, Shulevski et al. 2012, Brienza et al 2018). These large-scale lobes emerge from NGC 1167 and rise, interacting with the large-scale plasma and magnetic fields (Adebahr et al 2018), and are explained as a past phase of radio emission. The extended lobes contribute only 20% of the total radio luminosity, with a luminosity at 1.4 GHz of 5.5 x $10^{22}$ W Hz$^{-1}$ (Shulevski et al. 2012). Thus, B2 0258+35 shows multiple phases of activity, with the most recent one being a kpc-scale CSS: this young radio source and its impact on the surrounding medium is the focus of this study.

B2 0258+35 / NGC 1167 uniquely fills an unexplored regime; an earlier stage of a young mid-intensity jet in a cold ISM-rich early-type galaxy. This young jet is nested within the host galaxy and has been proposed to interact with the gas in the circumnuclear region. The scenario that has emerged for B2 0258+35 is that of jets impacting the surrounding gas and affecting multiple phases of the ISM. Observations of the cold gas (HI and molecular) in the circumnuclear regions of NGC 1167 have revealed irregular gas kinematics that can be attributed to the effect of the radio jet. By studying HI in absorption against the radio source and combining it with single-dish CO studies from the literature, Murthy et al. (2019) found that highly turbulent (velocity dispersion of ∼90 km s$^{-1}$) cold gas is concentrated in the central few kpc. By comparing with the numerical simulations of Mukherjee et al. (2018), Murthy et al. suggest that this turbulence is due to the jets expanding into a circumnuclear disc. Adding to this picture, a fast (∼500 km s$^{-1}$), massive (∼ $10^7$ M$_\odot$) molecular outflow has been imaged with the NOrthern Extended Millimeter Array (NOEMA) and is roughly located in the region of the sharp bend of the jet (Murthy et al. 2022). This outflow is carrying out about 75% of the gas in the central region, thus depleting the kiloparsec-scale molecular gas reservoir in a few million years.

If the above scenario is correct, the impact of the jet should also be seen in the hot gas traced by X-rays. The ∼$10^6$ yr age of the young, kiloparsec-sized, B2 0258+35 radio source (Giroletti et al. 2005) suggests that the jet has an expansion speed of a few 1000 km s$^{-1}$. This high propagation speed should lead to X-ray emitting shocks. Shock-heated hot gas resulting from the impact of radio lobes on galactic scales has been reported in X-rays in a few objects with strong radio jets, in particular, Cen A and NGC 3801 by Croston et al. (2007, 2009). These authors have characterized the properties of the X-ray gas and of the shocks produced by the expansion of the radio lobes. They have concluded that they can be energetically capable of affecting the evolution of the host (see also Webster et al. 2021). Signatures of this impact have also been seen in Seyfert galaxies, where Chandra observations have revealed shocked regions marking the interaction of weak jets with the dense cold ISM of the host spiral galaxy (e.g., Wang et al. 2011b; Fabbiano et al. 2018b). At L (1.4 GHz), = $2.1 \times 10^{23}$ W Hz$^{-1}$ (Giroletti et al. 2005) the B2 0258+35 jet is some ten times more powerful than that in NGC 4151 ($1.3 \times 10^{22}$ W Hz$^{-1}$ at 1.4 GHz, Johnston et al. 1982) and so the interaction should have a definite X-ray signature.

In this paper we investigate the resolved properties of the X-ray emission of the ionized gas in the central kpc-region of the NGC1167, co-spatial with the young radio jet. Moreover, we explore the physical state of the central engine, which has been suggested to be a Compton Thick (CT) AGN. NGC 1167 was included in an XMM-Newton spectral study of Seyfert 2 galaxies (Akylas & Georgantopoulos 2009), although as discussed above there is no evidence of Seyfert activity in the optical spectra (Ho et al. 2009; Gomes et al. 2016). Akylas and Georgantopoulos stack the X-ray spectrum of NGC 1167 with those of absorbed Seyfert 2 galaxies in their sample, and the stacked spectrum shows the 6.4 keV Fe Kα fluorescent line typical of CT AGNs. However, they do not detect this line in the spectrum of NGC 1167, leaving the nature of this AGN uncertain. If the nucleus of NGC 1167 were a CT AGN, we would expect it to be a relatively bright source in the hard X-ray band above 10 keV. As part of the present study, we have complemented the Chandra data with a joint NuSTAR 50 ks exposure to further constrain the hard X-ray emission, which we also discuss in this paper.

This paper is structured as follows: in Section 2, we discuss the Chandra observations and data analysis and present the results of this analysis; in Section 3, we discuss the NuSTAR observations, analysis and results; in Section 4 we discuss the astrophysical implications of our results; and, in Section 5 we summarize our conclusions.

## 2. Chandra Observations, Analysis and Results

Table 1 summarizes the Chandra observations of NGC 1167 used in this paper. Most of the data were obtained in our 2020 observing campaign. The first data set (PI: Goulding) was never published, but it was used by our team to design our deep observations. Together with the 2016 archival dataset, the total ACIS-S exposure time is 210.6 ks. The backilluminated S3 chip was used in all cases (see Chandra Proposer's Observatory Guide Section 6.1[1]). After filtering for time intervals of high background flux exceeding 3 σ the average level with the DEFLARE task, the resulting final exposure is 210.1 ks. All the data sets were processed to enable subpixel analysis (see Wang et al. 2011a, Fabbiano & Elvis 2022). These data have been analyzed with the CIAO (Fruscione et al. 2006) data analysis system version 4.13 and the Chandra calibration database CALDB version 4.9.6, adopting standard procedures; additionally, we made use of the Sherpa application (Freeman et al. 2001) for the spectral analysis.

### 2.1 Merged image and astrometry correction

We merged the observations to produce a deeper data set for imaging analysis, following the CIAO threads[2], and using the longest observation (ObsID 24860) as the astrometric reference. Images binned at 1/4 of the ACIS instrument pixel of 0.492" were used for the merging. We used both the centroid of the 0.3-7.0 keV emission in the nuclear region of NGC 1167 and external sources as reference points, with consistent results.

We then registered the merged image aligning the centroid of the hard 3.0 -7.0 keV band emission with a Pan-STARRS source located at RA=03:01:42.33 Dec=+35:12:20.3 (J2000), coincident with the core of the radio emission as observed at 22 GHz (Giroletti et al. 2005). This resulted in a shift of 0.506'', compatible with the ~1" uncertainty of the absolute astrometry of Chandra (POG Section 5.4.1 Celestial Location Accuracy[3]). We chose the hard-band centroid, because these energies are less affected by absorption in the nuclear region, and typically give a more precise position of the nuclear source. A single X-ray source is visible in this spectral band, which we assume as the X-ray AGN emission.

The final merged image in the 0.3-7.0 keV energy range, with ⅛ binning, is shown in the left panel of Figure 1. Within a radius of 3'' from the hard band peak, we detect 446 +/- 22 counts over the field background in the 0.3 – 7.0 keV band. The emission is mostly soft: only ~60 +/- 8 counts are detected in the same circle at energies >3.0 keV. The emission is also spatially extended, as can be seen in the right panel of Figure 1, which shows the 0.3 - 7.0 keV radial surface brightness profile, compared to that of the Chandra PSF in the same energy band. The energy and azimuthal dependences of the X-ray emission surface brightness are discussed in Section 2.2.

---

Table 1.- Chandra observations summary

| ObsID | Instrument | T$_{exp}$ (ks) | PI | Date |
|---|---|---|---|---|
| 19313 | ACIS-S | 12.9 | Goulding | 2016-11-25 |
| 23677 | ACIS-S | 20.8 | Fabbiano | 2020-11-13 |
| 24237 | ACIS-S | 16.9 | Fabbiano | 2020-11-14 |
| 24238 | ACIS-S | 16.9 | Fabbiano | 2020-11-09 |
| 24239 | ACIS-S | 23.3 | Fabbiano | 2020-11-11 |
| 24240 | ACIS-S | 18.1 | Fabbiano | 2020-11-16 |
| 24241 | ACIS-S | 23.6 | Fabbiano | 2020-11-19 |
| 24860 | ACIS-S | 26.7 | Fabbiano | 2020-12-14 |
| 24866 | ACIS-S | 15.9 | Fabbiano | 2020-11-16 |
| 24900 | ACIS-S | 20.7 | Fabbiano | 2020-12-18 |
| 24901 | ACIS-S | 14.8 | Fabbiano | 2020-12-19 |

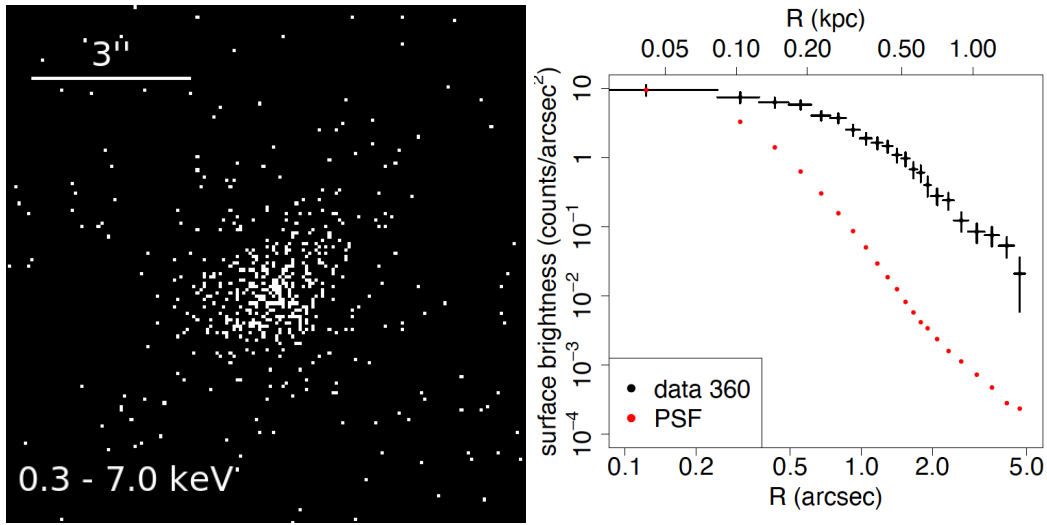

Figure 1 - Left: Observed 0.3 - 7.0 keV data with ⅛ binning. Right: The 0.3 -7.0 keV radial surface brightness profile centered on the position of the hard 3.0 -7.0 keV nuclear source (see text), compared with that of the Chandra Point Response Function (PSF) in the 0.3 - 7.0 keV band (red points) for the full 360° sector.

2.2 Comparison of the X-ray surface brightness distribution with the radio jet emission

Figure 2 compares the X-ray emission in three energy bands: total (0.3 – 7.0 keV), soft (0.3 – 3.0 keV) and hard (3.0 – 7.0 keV), from the images obtained by merging the individual observations, as described in Section 2.1. The images are displayed with 1/8 binning and are smoothed with a two-dimensional Gaussian filter of two image pixels. These images show that the peak of the X-ray emission is displaced by 0.26'' between the soft and hard images. Based on previous work on the Chandra images of nearby AGNs (e.g., ESO 420-G014, Fabbiano et al. 2018b) this offset may be indicative of obscuration of the nuclear source in the soft band.

Overplotted in Figure 2 are iso-density contours of radio emission at 22 GHz. In all cases the contours start from 2 mJy/beam and follow with power of two increases. Comparison of the left and central panels of Fig. 2 shows that the extended X-ray emission is dominated by the soft 0.3-3.0 keV emission.

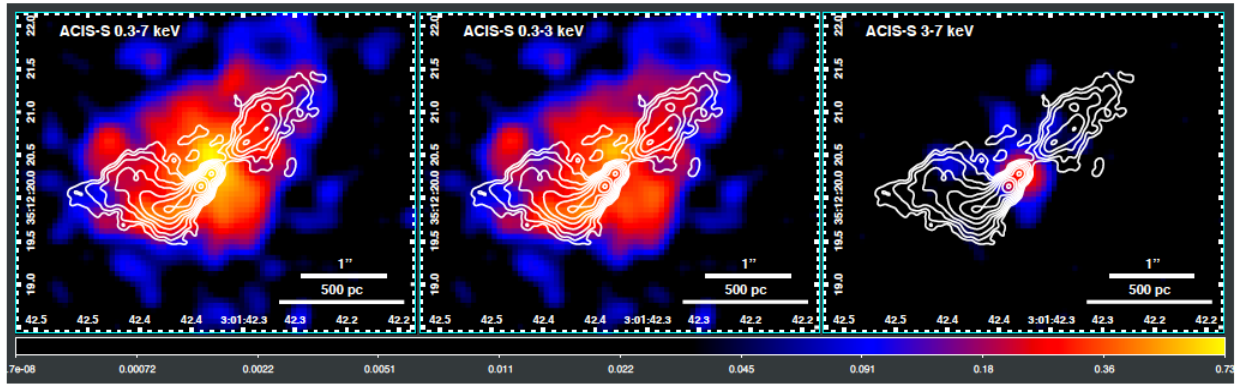

Figure 2 – Merged ACIS images of the central region of NGC1167 in different energy bands, with overlaid 22 GHz radio contours (Giroletti et al. 2005). The images are presented with a subpixel of 1/8 of the native ACIS pixel size and smoothed with a Gaussian kernel of σ = 2 image pixels.

Figure 3 shows the binned surface brightness profiles averaged in the two 90 deg. quadrants enclosing the jets and in the perpendicular direction, for the soft 0.3 - 3.0 keV band (top) and the hard 3.0 -7.0 keV band (bottom). The Chandra PSF profiles are also shown for the same energy bands, normalized to the central bin. As can be seen in Figure 3, the soft surface brightness profiles are clearly extended in both jet and cross-jet directions. The hard emission is also slightly extended, especially in the direction of the radio jet. While extended soft X-ray emission is common in AGNs (Bianchi et al. 2006; Levenson et al. 2006), extended hard emission has also been found with Chandra in several CT AGNs (e.g., ESO 428-G014, Fabbiano et al. 2017; Jones et al. 2021).

The azimuthal variations of the soft X-ray surface brightness are shown quantitatively by the azimuthal surface brightness profile in Figure 4. While there is extended emission at all azimuthal angles (in agreement with the radial profiles of Figure 3), the soft X-ray surface brightness is clearly enhanced in the direction of the radio jet, with both NW (angle 30-60 deg.) and SE (angle 180-240 deg.) extensions.

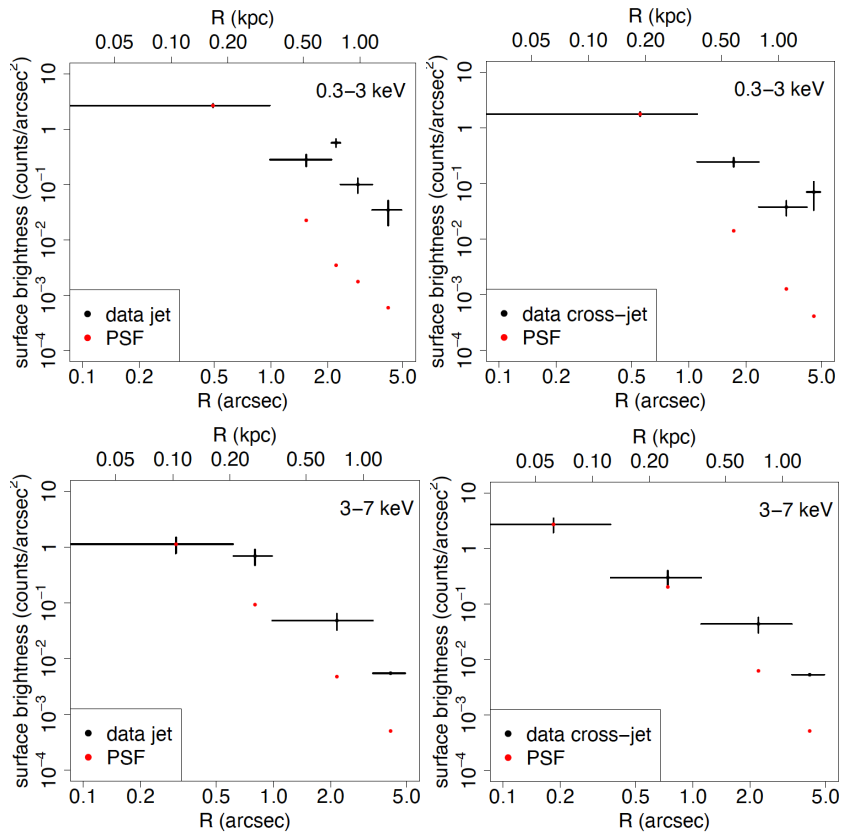

Figure 3 - Top: Radial profiles of the 0.3 -3.0 keV surface brightness, in two 90 deg. quadrants in the jet direction (top left), and in the cross-jet direction (top right). The PSF is indicated by red dots. Bottom: the same for the 3.0-7.0 keV band.

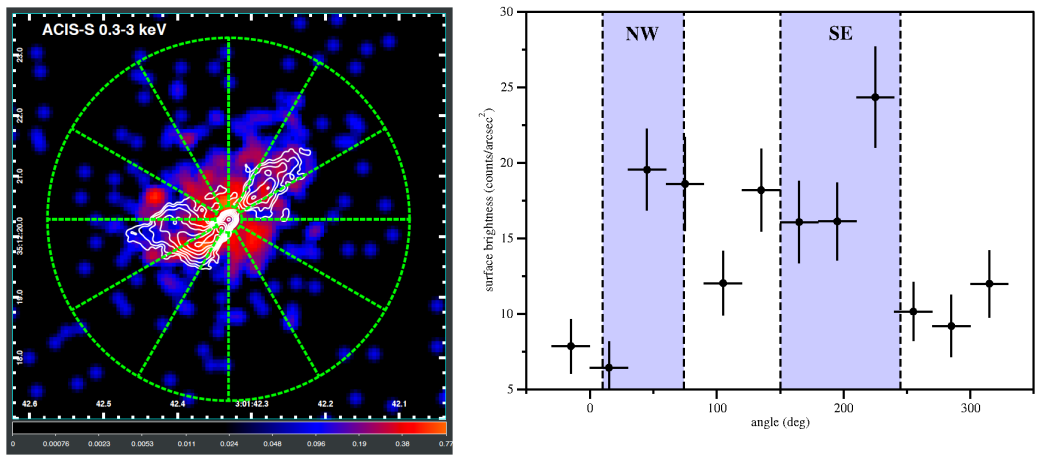

Figure 4 - Left - The 12 azimuthal sectors used to extract the average surface brightness. Right - azimuthal surface brightness profile, with bins starting from W and going counterclockwise. The jet emission regions are in the colored areas between the two sets of vertical lines.

To investigate spatial variations of the spectral properties, we have produced an X-ray Hardness Ratio (HR) map, shown in Figure 5. HR = (H-S) / (H+S), where H indicates the counts detected in the 2.0 - 7.0 keV band, while S indicates those in the 0.3 -2.0 keV band, respectively. The choice of the 2.0 keV cut was based on the overall count spectral distribution, to isolate the spectral region dominated by the soft emission lines (see Section 2.3 and Figure 7). The ratio map has been produced using band images with a subpixel of ¼, and then smoothed with a Gaussian kernel with a σ = 2 image pixels.

The left panel of Figure 5 shows the 22 GHz contours superimposed on the HR map. Most of the extended X-ray emission has a similar 'harder' HR (green areas in Figure 5). The regions coinciding with the radio core and inner jet are also hard. Instead, the region where the SE side of the jet bends and the region immediately to the South of the bend are co-spatial with the softer X-ray emission (blue pixels). The right panel of Figure 5 shows the comparison of the X-ray surface brightness distribution and the HR. This figure illustrates that the region of softer emission in the HR map is limited to the southern region of the distribution of X-ray emission. As demonstrated by the spectral analysis (Section 2.2), this region is fitted with a significantly lower kT than the surrounding higher HR area (~1 keV versus ~2.2 keV, Table 4).

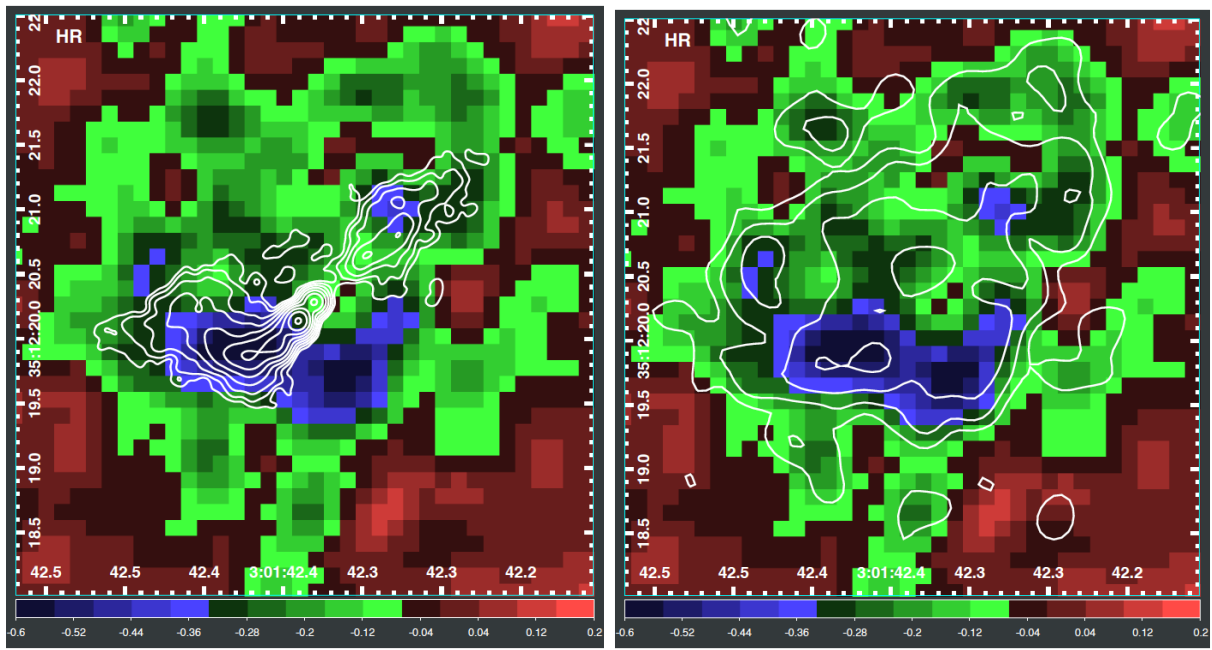

Fig. 5 - Left: Hardness Ratio (HR) map (see text for definition), with 22 GHz contours superimposed. N to the top and E to the left. Note the lower HR pixels (blue) along the jet and to the south of the bend of the jet. Right: HR map with 0.3 - 3.0 keV iso-intensity contours from the X-ray image in Fig. 2. The soft (blue pixel) region to the south of the radio jet is a region of intense X-ray emission.

2. 3 Spectral analysis of the Chandra data

Spectra were extracted in the regions presented in Figure 6 (left), namely the north (N) and south (S) regions on the sides of the radio jet, the radio core/nucleus, and the radio jet (excluding the core region, but including both jet and counter-jet, see Figure 6), and from a circular region with a 4'' radius centered on the radio core location, encompassing the entire extended emission. The background was extracted from a larger region outside the extended source. We also extracted spectra from two additional regions, based on the HR map, as shown in Figure 6 (right): the region encompassing the soft HR values (blue pixels) and the encompassing harder HR values region (green pixels).

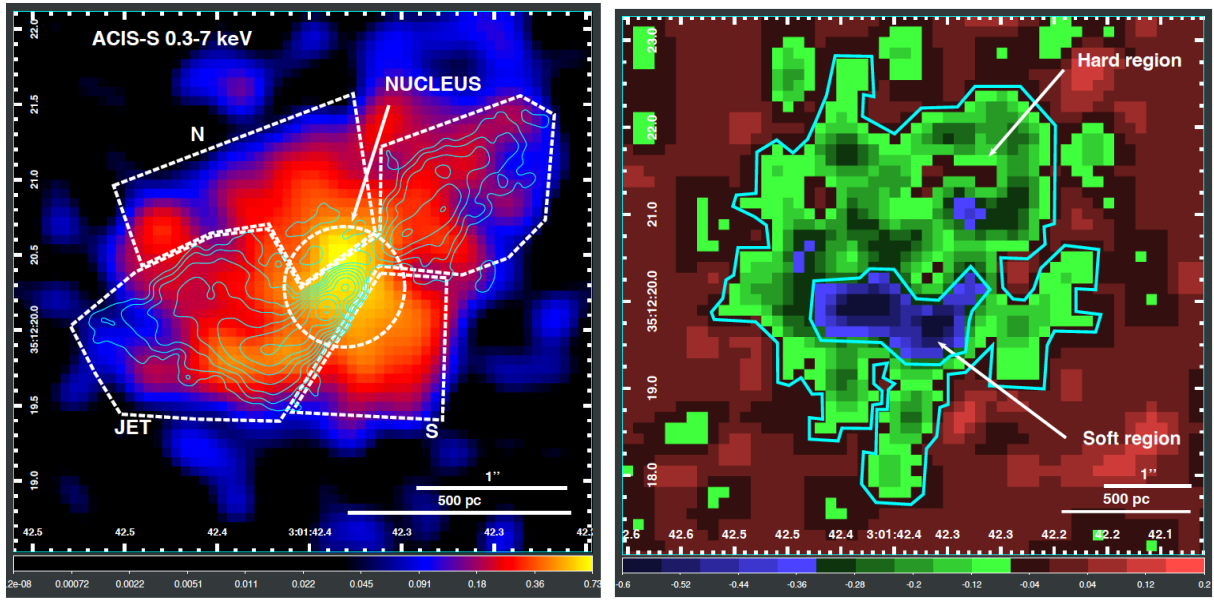

Figure 6 - Left: 0.3 -7.0 KeV image with the 22 GHz radio contours are overplotted in cyan. The regions used for spectral extraction are shown with white dashed lines. The 'All' region is a circle of 4'' radius, encompassing the entire X-ray emission. Right: HR image with extraction regions. Note the different spatial scales.

Due to the low count statistics, instead of subtracting the background spectra we modeled them using the prescription given by Markevitch et al. (2003), that is, a model comprising a thermal plasma component (MEKAL, Kaastra 1992) with solar abundances, and a power law. Given the small number of detected counts, for this analysis we binned the spectra to obtain a minimum of 1 count per bin, making use of the Cash statistic. For spectra with >120 counts, we have also used the $\chi^2$ statistic, binning the data in 20 counts/bin. The uncertainties on all the spectral parameters reported below are 1 σ for that interesting parameter.

The spectrum from the entire X-ray emission region ('All', with 463 net counts) suggests the presence of emission lines. To investigate this possibility, we fitted this spectrum with a power-law model, with power-law index fixed to 1.8, incrementally adding gaussian lines to the model, leaving both line energy and amplitude free to vary (see Fabbiano et al. 2018a for a detailed description of this procedure). The spectrum and the best-fit model are shown in Figure 7.

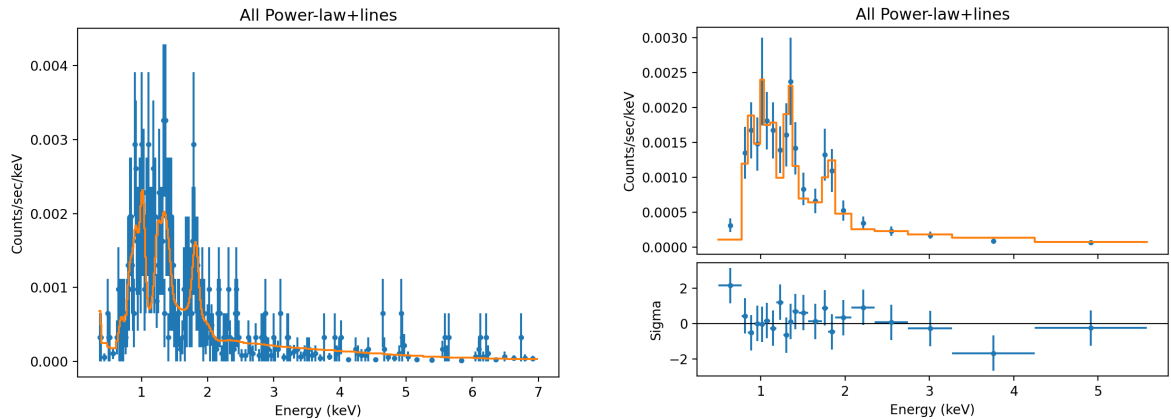

Figure 7 - Spectrum and best fit model for the region encompassing the entire diffuse emission (the unbinned spectrum and best-fit model using Cash statistics, on the left; the binned spectrum and the $\chi^2$ best fit model and residuals on the right.)

This fit identifies possible emission from lines (Tables 2, 3), which are often found in AGN ACIS spectra of significantly larger S/N (e.g., Wang et al. 2011c; Fabbiano et al. 2018a; Travascio et al. 2021; Paggi et al. 2022). In particular, both Cash and $\chi^2$ fits identify a significant Mg XI (1.352 keV) and possible Si XIII (1.839 keV) emission. Also, Ne X (1.022 keV) emission is found in both cases, while the Cash fit suggests in addition Ne IX (0.915 keV). Both Ne lines have been associated with shocked ISM emission (Wang et al. 2011; Paggi et al. 2012). Different transitions of Fe XIX are suggested in the two fits. Other lines are suggested in the different fits, possibly as a result of the relatively poor statistics and the line blending due to the limited spectral resolution of ACIS.

Table 2. Power-Law + Lines fit to 'All' Spectrum - Cash

| Best-Fit Lines | | |
|---|---|---|
| Line | Rest-frame energy (keV) | Line Flux ($10^{-5}$ photons cm$^{-2}$ s$^{-1}$) |
| Fe XVIII | 0.704 | 0.38 (-0.16+0.20) |
| Fe XVII | 0.826 | 0.26 (-0.09+0.10) |
| Ne IX | 0.915 | 0.20 (-0.06+0.06) |
| Ne X | 1.022 | 0.16 (-0.03+0.04) |
| Fe XIX $1s^22s^22p^4 \rightarrow 1s^22s^22p^35d^1$ | 1.258 | 0.04 (-0.02+0.02) |
| Mg XI | 1.352 | 0.04 (-0.01+0.02) |
| Fe XXII | 1.423 | 0.02 (-0.01+0.01) |
| Si XIII | 1.839 | 0.02 (-0.01+0.01) |
| **Other Fit Results** | | |
| PL Norm. for ($10^{-5}$ keV$^{-1}$ cm$^{-2}$ s$^{-1}$) | C (d.o.f.) | $F_{0.3-7\ keV}$ ($10^{-14}$ cgs) |
| 0.40 (-0.03+0.03) | 1.05 (634) | 3.83 (-0.33+0.26) |

Table 3. Power-Law + Lines Fit to 'All' Spectrum - $\chi^2$

| Best-Fit Lines | | |
|---|---|---|
| Line | Rest-frame energy (keV) | Line Flux ($10^{-5}$ photons cm$^{-2}$ s$^{-1}$) |
| O VIII | 0.871 | 0.42 (-0.09+0.09) |
| Ne X | 1.022 | 0.18 (-0.04+0.04) |
| Fe XIX $1s^2 2s^2 2p^4 \rightarrow 1s^2 2s^2 2p^3 4d^1$ | 1.146 | 0.08 (-0.02+0.02) |
| Mg XI | 1.352 | 0.06 (-0.02+0.02) |
| Si XIII | 1.839 | 0.02 (-0.01+0.01) |
| **Other Fit Results** | | |
| PL Norm. ($10^{-5}$ keV$^{-1}$ cm$^{-2}$ s$^{-1}$) | $\chi^2$(d.o.f.) | $F_{0.3-7 \text{ keV}}$ ($10^{-14}$ cgs) |
| 0.37 (-0.04+0.04) | 0.85 (15) | 3.22 (-0.16+0.18) |

With the exception of the spectrum from the nuclear region, which contains only 65 counts, we followed our previous work (e.g., Fabbiano et al. 2018a), and fitted the spectra with two different physical models, which would give rise to line emission in the spectra: a simple thermal plasma model (APEC) with solar abundances, and a photo-ionization model (CLOUDY). For the nuclear region, instead, we used a simple power-law model. In all the models we have included photo-electric absorption by the Galactic column density along the line-of-sight $N_H = 9.68 \times 10^{20}$ cm$^{-2}$ (HI4PI Collaboration et al. 2016). We used Cash statistics in all cases, and we also performed $\chi^2$ fits for the spectra with >120 counts. A second spectral component was added if the $\chi^2$ could be improved. The results are summarized in Tables 4, 5, and 6. Table 6 gives the derived physical gas parameters, from the C-stat best-fit values. The log ($N_H$) values marked with an asterisk in Table 6 are the best-fits from fits in which this parameter could not be constrained, and was therefore frozen to the best fit value.

In most cases the fits are acceptable, which is perhaps not surprising given the low signal to noise of these spectra. However, the 'Jet' and 'All' spectra may require more complex models, at least for the thermal fits. The kT in all the cases (except for the 'Soft' region) is > 1.5 keV. The 'Soft' region of the HR map returns significantly lower temperature (kT ~ 1.0 keV) than the entire 'Hard' extended emission area (kT ~ 2.25 keV) surrounding it. The spectra are generally reasonably well fitted by the photoionization models.

Figure 8 shows the 'All' spectrum fitted with a 2-component thermal model (on the left) and with a photoionization model (on the right). In both cases the fits are good (see Tables 4 and 6), but there are areas of correlated residuals suggesting perhaps a more complex situation, where both photoionization and shock ionization may occur. These complex spectra have been detected in other AGNs observed with much better statistics (e.g., ESO 428 - G014, Fabbiano et al. 2018a; NGC 6240, Paggi et al. 2022).

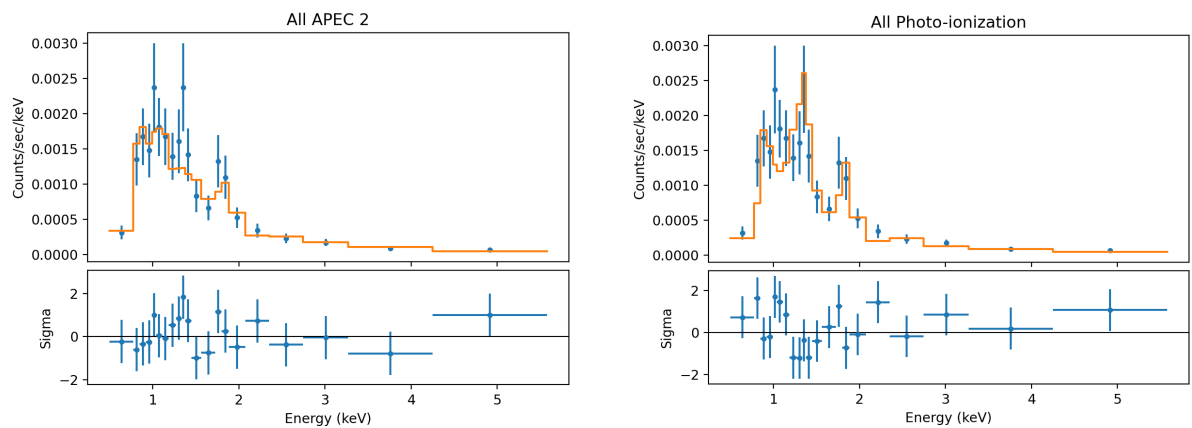

Figure 8 - Binned spectrum of the entire X-ray emission extracted from a circle of 4" radius centered on the nucleus, with $\chi^2$ best fit model and residuals. Left: best-fit APEC model in orange. Right: Best-fit CLOUDY model in orange. In both cases the fit residuals are displayed in the lower panels.

Given the limited statistics, not much can be learned from the spectrum of the nuclear region (Figure 9, Table 7), except that the distribution of counts may suggest a grouping consistent with emission lines for energies <2.5 keV. More interestingly, there is no evidence of a prominent Fe I Kα line at 6.4 keV, which would be indicative of a Compton Thick AGN. By fitting the continuum between 4-6 keV, from this spectrum we derive an equivalent width EW = 530 (-400, +820) eV for this line (where the uncertainties are at the 1 σ level,) to be compared with a typical EW~1000 eV for a CT AGN. Given the poor statistics we cannot exclude a CT AGN, although the spectrum is also consistent with no 6.4 keV line. However, the non-detection of this source with NuSTAR (see Section 3) confirms that NGC 1167 does not host a CT AGN. The nuclear X-ray flux (Table 7) corresponds to a 0.3-7.0 keV luminosity of the nuclear source is ~4 x $10^{39}$ erg s$^{-1}$, for the distance of 70 Mpc.

Table 4. Thermal Fits

| Region | Net Counts (0.3-7.0 keV) | kT | C (d.o.f.) | $\chi^2$(d.o.f.) | $F_{0.3\text{-}7\,keV}$ ($10^{-14}$ cgs) |
|---|---|---|---|---|---|
| N | 75 | 1.94 (-0.52+0.53) | 1.09 (501) | | 0.51 (-0.12+0.15) |
| S | 55 | 2.33 (-0.53+0.74) | 1.07 (497) | | 0.37 (-0.05+0.07) |
| N+S | 129 | 2.04 (-0.19+0.68) | 1.14 (533) | | 0.86 (-0.11+0.12) |
| | | 2.25 (-0.44+0.90) | | 2.22 (3) | 0.70 (-0.09+0.12) |
| Jet | 163 | 1.67 (-0.25+0.25) | 1.09 (556) | | 1.01 (-0.10+0.16) |
| | | >2.42<br>0.65 (-0.25+0.38) | | 0.85 (3) | 1.56 (-0.52+0.36) |
| All | 463 | 1.59 (-0.11+0.21) | 1.17 (641) | | 2.79 (-0.21+0.15) |
| | | 3.40 (-0.77+1.41)<br>0.65 (-0.14+0.10) | | 0.70 (17) | 3.88 (-0.35+0.41) |
| Soft | 120 | 0.99 (-0.12+0.09) | 1.11 (498) | | 0.87 (-0.11+0.15) |
| | | 1.61(-0.22+0.35) | | 0.93(2) | 0.66 (-0.11+0.17) |
| Hard | 270 | 2.25(-0.21+0.33) | 1.13(594) | | 1.77 (-0.13+0.10) |
| | | 3.75(-1.09+4.85)<br>0.73(-0.52+0.22) | | 0.36(8) | 1.16 (-0.20+0.22) |

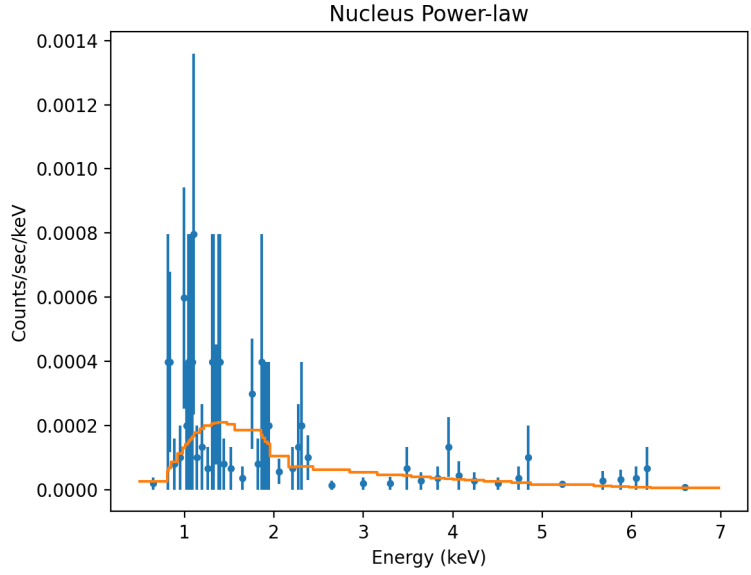

Figure 9 - Unbinned spectrum from the nucleus with best-fit Cash power-law model in orange.

Table 5. Physical parameters from thermal Fits

| Region | Density (10⁻²⁵ g/cm³) | Pressure (10⁻⁹ Ba) | Cooling time (10⁶ yr) |
|---|---|---|---|
| N | 5.34 (-0.51+0.42) | 1.55 (-0.53+0.73) | 36.7(-13.5+15.0) |
| S | 7.22 (-0.51+0.52) | 2.60 (-0.77+1.01) | 33.1(-9.1+12.5) |
| N+S | 6.05 (-0.27+0.29) | 2.10 (-0.41+0.59) | 36.4(-6.0+13.3) |
| Jet | 5.79 (-0.49+0.29) | 1.49 (-0.39+0.29) | 29.2(-5.8+6.5) |
| All | 0.87 (-0.03+0.04) | 0.21 (-0.02+0.04) | 180.0(-19.4+27.0) |
| Soft | 4.84 (-0.23+0.34) | 0.74 (-0.08+0.19) | 12.5(-2.3+2.6) |
| Hard | 1.80 (-0.06+0.06) | 0.63 (-0.08+0.11) | 132.7(-16.4+21.3) |

Table 6. Photoionization Fits

| Region | Net Counts (0.3-7.0 keV) | Log (U) | Log ($N_H$) | C (d.o.f.) | $\chi^2$(d.o.f.) | $F_{0.3-7\text{ keV}}$ ($10^{-14}$ cgs) |
|---|---|---|---|---|---|---|
| N | 75 | 1.21 (-0.14+0.13) | 22.5* | 1.04 (501) | | 0.68 (-0.19+0.25) |
| S | 55 | 0.86 (-0.15+0.22) | 22.0* | 1.02 (497) | | 0.65 (-0.33+0.52) |
| N+S | 129 | 1.08 (-0.06+0.11) | 22.4* | 1.05 (533) | | 1.28 (-0.38+0.45) |
| | | 0.50 (-0.24+0.40) | 21.86 (-0.57+1.04) | | 1.40 (2) | 2.74-1.74+3.59 |
| Jet | 163 | 1.04 (-0.21+0.09) | 21.7* | 1.02 (556) | | 1.65 (-0.43+0.47) |
| | | -0.09 (-0.15+0.28) | 20.30 (-0.76+0.83) | | 1.22 (4) | 6.63 (-4.41+10.39) |
| All | 463 | 1.10 (-0.05+0.06) | 22.1* | 1.03 (641) | | 4.26 (-0.70+0.67) |
| | | 1.07 (-0.05+0.08) | 22.06 (-0.34+0.33) | | 1.12 (18) | 3.25 (-1.55+3.54) |

Table 7. Nucleus – Power Law Fit

| Net Counts (0.3-7.0 keV) | $\Gamma$ | C(d.o.f.) | $F_{0.3-7\text{ keV}}$ ($10^{-14}$ cgs) |
|---|---|---|---|
| 65 | 1.85 (-0.25+0.25) | 0.99 (480) | 0.62 (-0.13+0.15) |

### 3. NuSTAR Observations, Analysis and Results

The target was observed with NuSTAR (Harrison et al. 2013) starting on UT 2020 December 24 (OBSID 50701001002). The raw data, provided through HEASARC, were processed using HEASOFT version 6.28 and NuSTARDAS version 2.0.0 with CALDB version 20200912. We performed event filtering with the help of the *nupipeline* script with the following settings: *saacalc=3 saamode=optimized tentacle=yes*. The resulting total exposure is 48.9 ks. We find no significant detection (>3 $\sigma$) at the position of the NGC 1167 nucleus, neither in the softer 3-8 keV, nor in the harder 8-24 keV bands. We estimate that the 3 $\sigma$ upper limits for the flux in FPMA, which has lower background at the position of the target, are $8 \times 10^{-14}$ erg/s/cm$^2$ and $2 \times 10^{-13}$ erg/s/cm$^2$ in the 3-8 keV and 8-24 keV bands, respectively. These correspond to upper limits on the observed X-ray luminosity of $4.7 \times 10^{40}$ erg s$^{-1}$ and $1.2 \times 10^{41}$ erg s$^{-1}$, respectively, to be compared with the Chandra (0.3 - 7 keV) measurement of $4 \times 10^{39}$ erg s$^{-1}$ (Table 7).

Lack of a significant detection implies that NGC 1167 does not harbor a Compton-thick AGN of a significantly higher intrinsic luminosity than that implied from its nuclear soft X-ray flux, although such a statement is somewhat model-dependent. To further test this, we assume an X-ray spectrum consistent with nearby Seyferts obscured in the line of sight by CT column density (e.g., Marchesi et al. 2019, Torres-Albà et al. 2021), so that the soft X-ray power law continuum observed at 3-7 keV with Chandra is due to Thomson-scattered light at the level of a few percent of the intrinsic coronal continuum (e.g., Gupta et al. 2021). For both continua we use a power-law of photon index $\Gamma = 1.8$ in two possible scenarios described below.

In a transmission-dominated case, where any contribution from reprocessing in the obscuring torus is neglected, we find that a nucleus with an intrinsic 2-10 keV luminosity of about $5 \times 10^{41}$ erg s$^{-1}$ would be significantly detected in the 8-24 keV band for $N_H < 2 \times 10^{24}$ cm$^{-2}$, where $N_H$ is the line-of-sight column density. In a scenario where reprocessing in the torus dominates the spectrum at >7 keV (while the intrinsic continuum is obscured with $N_H > 3 \times 10^{24}$ cm$^{-2}$ and subdominant in the hard X-ray band; e.g., Baloković et al. 2014, Buchner et al. 2021), the nucleus would have been significantly detected in both NuSTAR sub-bands if its intrinsic luminosity in the 2-10 keV band exceeded approximately $3 \times 10^{41}$ erg/s. These estimates have a systematic uncertainty of at least a factor of two given the range of possible spectral models and reasonable ranges of spectral parameters such as the torus covering factor (e.g., Baloković et al. 2018, 2021).

### 4. Discussion

In the previous Sections we have reported the results of the X-ray observations of NGC 1167 with Chandra and NuSTAR, and we have compared these results with radio observations. NGC 1167 is an SA0 galaxy that is unusually rich in warm and cold ISM with disturbed kinematics proposed to be likely from jet-ISM interaction (O'Sullivan et al. 2015; Gomes et al. 2016; Murthy et al. 2022). NGC 1167 hosts the low-luminosity, young (~0.4 Myr), Compact Steep Spectrum radio source B2 0258+35 (Giroletti et al. 2005, Shulevski et al. 2012). This radio source has a 1-kpc-size core-double-jet morphology (Brienza et al. 2018) and is embedded in the ISM of the host galaxy. Below we discuss the implications of our results for the X-ray properties and activity of NGC 1167 (Section 4.1), suggesting that both photoionization and thermal components may be present. We look at the implications of our results for the radio-jet host galaxy interaction in Section 4.2: we first discuss the evidence of large-scale shocks in NGC 1167 and of the compression of the hot ISM by CO outflows (Murphy et al. 2022) in the region where the southern jet terminates; we then discuss the large-scale morphology, which suggests lateral outflows, and compare it with hydrodynamic simulations tailored to NGC 1167/B2 0258+35. We further discuss our results in the light of emission models of young radio sources (Section 4.3) and compare the CO and X-ray distributions for possible evidence of X-ray induced CO depletion (Section 4.4). We conclude by discussing evidence that points to NGC 1167 as a 'turned off' AGN (Section 4.5) and suggesting a radiatively inefficient accretion flow (RIAF; Yuan and Narayan 2014) state for its nucleus (Section 4.6).

## 4.1 The Extended X-ray Emission of NGC 1167

The morphology of the X-ray emission of NGC 1167 is reminiscent of that of other nearby radio-quiet AGN images obtained with Chandra, with an elongation of the X-ray surface brightness in the direction of the radio jet (Figures 2, 4), which typically also follows the emission line bicone (e.g., NGC 4151, Wang et al. 2011b; ESO 428-G014, Fabbiano et al. 2017; IC 5063, Travascio et al. 2021.) The elongation is less pronounced than in the radio-quiet AGNs (see e.g., review by Fabbiano & Elvis 2022) and NGC 1167 shows more prominent extended emission in the direction perpendicular to the radio jet than some other AGNs. However, a similar morphology has been observed in other cases (e.g., in ESO 428-G014 and IC 5063, see Fabbiano & Elvis 2022) and it may be caused by the jet-ISM interaction (see Section 4.3).

Similar to other AGNs studied with Chandra, there is some indication of extended emission in the hard band, at energies > 3.0 keV (Figure 3). At these energies, the spectra of obscured AGNs are dominated by a flat continuum and the fluorescent Fe Kα 6.4 keV line. Although our statistics are limited, the spectrum of the extended emission is consistent with these other observations of obscured AGNs. In the standard AGN model, of an active nucleus obscured by a nuclear torus, the high energy emission (>3 keV) is believed to be due to nuclear photons reflected and reprocessed by the obscuring torus and should therefore be point-like in Chandra observations. This paradigm has been at least partially falsified by the detection of extended X-ray components at energies > 3.0 keV, which suggest X-ray reprocessing in dense molecular clouds in the host galaxy disk (e.g., Fabbiano et al. 2017, 2018b; Jones et al. 2021; see review Fabbiano & Elvis 2022). NGC 1167 is rich in cold ISM and molecular clouds (see Section 4.2); while the limited statistics do not allow us to determine if the surface brightness above 3 keV is elongated in the jet direction, the presence of extended hard emission is uncontroversial (Figure 3).

In some active heavily obscured AGNs with small radio jets (NGC 4151, Wang et al. 2011b; Mkn 573, Paggi et al. 2012; NGC 3393, Maksym et al. 2017), the extended X-ray emission along the jets has been found to be a mix of photoionization associated with the ionized bicones together with somewhat more localized thermal components that are more closely associated with the jet and its termination. The spectral data of NGC 1167 does not have enough signal-to-noise to firmly require multi-component fits.

In the photoionization model fits (Table 6), the best-fit ionization parameter (log U) values are in the range of those observed in other AGNs (see the compilation in Figure 5 of Travascio et al.

2021). Since photoionization is favored in nearby AGNs observed with better statistics (e.g., NGC 4151, Wang et al. 2011c), we cannot exclude this possibility for NGC 1167.

In the thermal model fits (Table 4), with the exception of the 'Soft' emission region, the values of kT are in the 1.5 - 3 keV range. Such high values are not consistent with a hot halo trapped in the dark matter potential of NGC 1167, since they exceed the typical temperatures of the hot halos of early type galaxies. Given the total X-ray luminosity of NGC 1167, $L_X \sim 2.5 \times 10^{40}$ erg s$^{-1}$ (for the total emitted flux in Table 4 and a distance of 70 Mpc), the observed relation between $L_X$ and kT of E and S0 galaxies predicts a kT < 1.0 keV (Kim & Fabbiano 2015).

The kT values >1.5 keV implied by the thermal fits require energy input from an AGN. Taken at face value, they may suggest strong shocks from outflows with velocities of >1000 km/s, as observed in the well-studied case of NGC 6240 (Feruglio et al. 2013; Wang et al. 2014). Are we seeing the X-ray emission from the 'shocked' cocoon around the jet, with enhancements at the hot spots (e.g., Heinz et al. 1998)? Shocks driven in the ISM by the impact of the jet may provide this additional energy input. In IC 5063, there is direct evidence of strong shocks in the jet-ISM interaction region because of localized Fe XXV emission (Travascio et al 2021), but in NGC 1167 we do not have the photon statistics necessary to resolve this emission, if present. Of course, it may simply be that in most regions the emission is dominated by photoionization, so that the thermal model alone does not apply.

The spectrum suggests line emission below 2 keV, with lines typically detected in other AGNs (see Tables 2 and 3, and review Fabbiano & Elvis 2022). While emission lines can be caused by either thermal or photoionized emission, the presence of Ne IX (0.915 keV) and Ne X (1.002 keV) lines suggests a component of shock ionization, that may result from the interaction of a radio jet with the ISM (e.g., in NGC4151, Wang et al. 2011b; Mkn 573, Paggi et al. 2012). In NGC4151 (Wang et al. 2011b) photoionization fits alone work well except in a region of enhanced NeIX/OVII. At this location an additional thermal (APEC) component produces a good fit. This enhanced NeIX/OVII region is coincident with the terminations of the weak radio jet in NGC 4151. A similar situation is seen in MRK 573 where both Ne IX and Ne X enhancements are seen, coincident with the termination of the radio jet (Paggi et al. 2012). We suggest then that Ne IX and Ne X may similarly result from shocks due to the interaction of the NGC 1167 radio jet with the ISM.

As shown in Section 2.2, there is also significant softer X-ray emission (kT ~ 1 keV, if thermal) in a region coincident with and to the south of the SE side of the radio jet (Section 2.1). This is the region where turbulence and a CO outflow has been recently detected (Murthy et al. 2020). We discuss the implications of this softer emission below (Section 4.2.1).

4.2 The Jet – ISM Interaction

The principal scientific objective of our Chandra observations was the high-resolution investigation of the jet/galaxy interaction in the X-rays. The new-born radio jet phase may be most effective in impacting the surrounding medium (see simulations of Wagner et al. 2012, Cielo et al. 2018). As discussed in Section 4.1, X-ray observations of nearby AGNs with disk-embedded jets have revealed thermal emission from the shocked ISM at the jet terminations (see Fabbiano & Elvis 2022). Numerical simulations (Mukherjee et al. 2018) also suggest that the expansion of low-luminosity radio jets can be significantly slowed down by the clumpy ISM in the central kpc, giving enough time for them to percolate through the disk. This can then drive shocks out to several tens of kpc in the galaxy, thereby contributing to negative feedback at distances much larger than the apparent physical reach of the jets. These simulations predict hot lateral outflows perpendicular to the galaxy disk and jet, which would translate into X-ray emitting gas to be found both in the jet direction (which in several AGNs would also contain the optical ionization cones, Fabbiano & Elvis 2022) and the cross-jet direction. Below we discuss these two effects in the case of NGC 1167/B2 0258+35.

4.2.1 Shock heated ISM and jet termination effects

As discussed in 4.1, the presence of Ne IX and Ne X spectral lines in the X-ray emission of NGC 1167 suggests the presence of some shock-ionization. Strong shocks in the central region of NGC 1167 are supported by its optical spectral characteristics. From the analysis of the IFU CALIFA data, Gomes et al. (2016) identify two regions in the disc of NGC 1167 that both show LINER spectra. Of these the inner region (radius ~4″; 1.4 kpc) is characterized by high equivalent-width (EW) Hα emission (~3 - 8 Å, see Figure 1c in Gomes et al. 2016), suggesting that evolved pAGB stars alone cannot be the source of ionization, as they can only generate an EW of Hα in the range 0.5 - 2.4 Å. Gomes et al. suggest that the more likely cause of the LINER spectra in the inner region is the presence of wide-spread shocks. The conclusions from the studies of cold gas in this source in combination with the numerical simulations of similar systems indicate that radio jets interacting with the ISM can very well give rise to such wide-spread shocks. The energetics of the radio jet in NGC 1167 is sufficient to produce this impact, including both the outflow and the high turbulence (see Murthy et al. 2019, 2022).

Murthy et al. (2022) discovered that the region where the southern side of the jet bends, is both coincident with the peak of the CO intensity and has higher velocity dispersion (see the right panel of Figure 10), suggesting the presence of turbulence and shocks, excited by the impact of the jet on the molecular clouds. In this region, located about 540 pc from the radio core, outflowing CO clouds have been observed, with a velocity of up to 500 km/s, (Murthy et al. 2022). This region, as also shown in Figure 10, approximately coincides with the area where the X-ray emission is significantly softer, with a cooler emission temperature in the thermal model fit (Figure 5, Table 4). Table 5 suggests that the density of the X-ray emitting gas may be ~ 2-3 times larger than in the surrounding harder emission region, and the resulting cooling time ~10 times lower. Taken at face value, these comparisons suggest that the ISM in the region of the jet bend may be compressed by the outflow detected in the molecular clouds observations, resulting in more efficient cooling. Also, the observed softer X-ray emission in this area argues against the presence of absorbing clouds in front of the X-ray emitting region, suggesting that the X-ray emitting gas is not co-located with the region containing the CO emitting clouds, but is external to it, residing between the CO clouds and the observer.

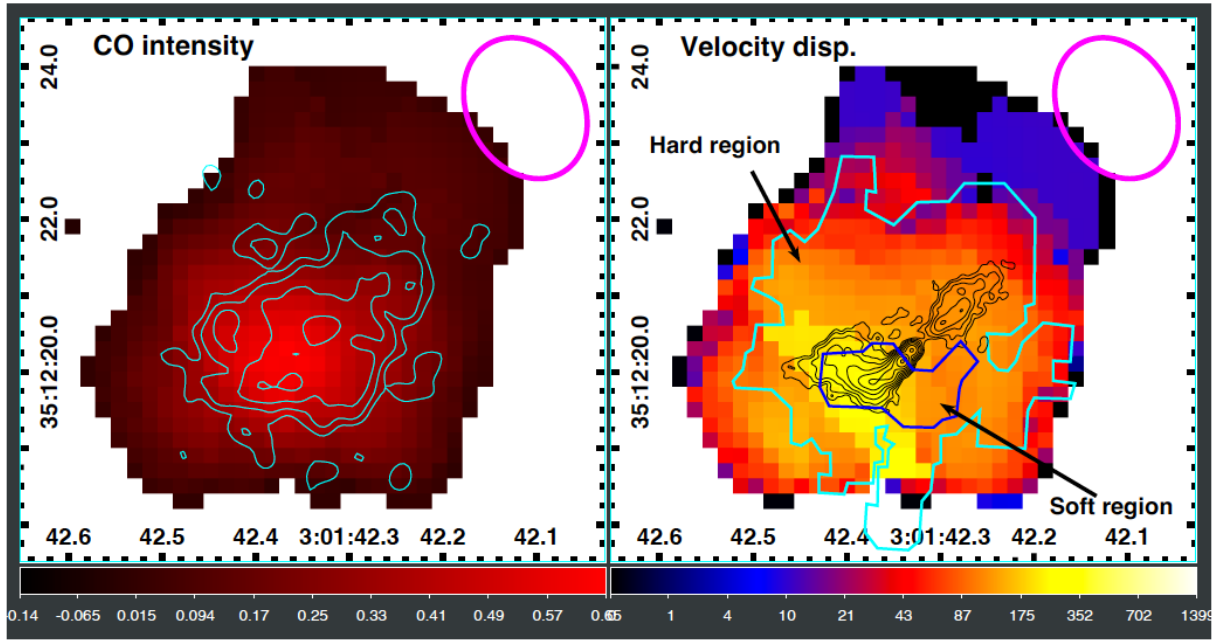

Figure 10 - Images from the CO observations presented in Murthy et al. (2022). Left: CO total intensity with superposed the Chandra full band (0.3-7 keV) contours derived from the 0.3 -3.0 keV image shown in Fig. 2. Right: CO velocity dispersion. The black contours are from the 22 GHz radio image (Giroletti et al. 2005); the blue and cyan contours identify the outer boundaries of the regions of soft and hard emission in the HR map. The magenta ellipses represent the beam of the CO observations.

### 4.2.2 Lateral outflows and jet-ISM interaction hydrodynamic simulations

Cross-jet extended emission is commonly found in nearby CT AGNs observed with Chandra, although it cannot be excluded that, at least in part, this emission may be due to a porous circumnuclear obscuring torus (e.g., ESO 428-G014, Fabbiano et al. 2018a, b; IC 5063, Travascio et al. 2021; see Fabbiano & Elvis 2022). In one case at least, the discovery of emission line filaments and a giant loop in the narrow band filter observations of IC 5063, presents convincing evidence of lateral outflows (Maksym et al. 2021). The Chandra observations of NGC 1167 / B2 0258+35 show that: (1) the soft (< 3.0 keV) X-ray emission is elongated in the direction of the radio jet and appears to envelop the jet (Figures 2, 11); and that (2) there is also prominent extended emission in the direction perpendicular to the jet (Figures 2, 3, 4, 11). Both could be consistent with the simulations of Mukherjee et al. (2018).

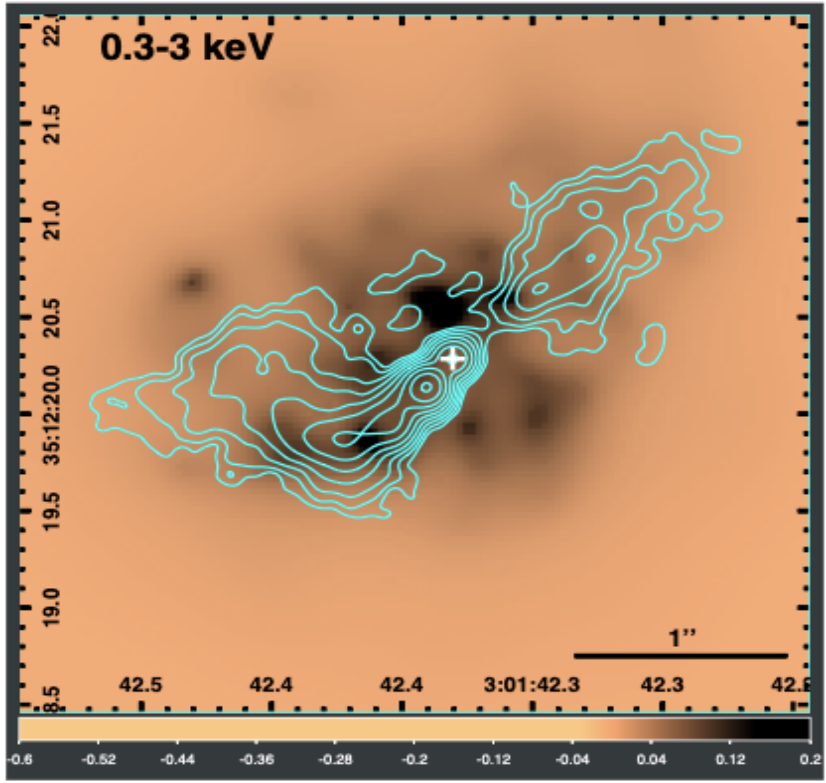

Figure 11. High contrast, adaptively smoothed image of the 0.3 - 3.0 keV X-ray surface brightness, with the 22 GHz radio contours. As in Fig. 2, the radio core position has been matched to that of the hard X-ray centroid (white cross).

Following Mukherjee et al. (2018), new hydrodynamic simulations were performed, which better demonstrate the qualitative similarities between the jet-ISM interactions seen in simulations to the case of NGC 1167. A jet of power $P_j \sim 10^{44}$ erg s$^{-1}$ was injected into a fractal gas disk of radial extent of ~1.8 kpc and mean density $n_{w0} = 100$ cm$^{-3}$. The power of the jet is an order of magnitude lower than in the previous simulations of Mukherjee et al. (2018), and hence is more appropriate for studying the driving of changes to the ISM of NGC 1167, which harbors a jet of comparable power (Murthy et al. 2022). The jet is launched at an angle of $45^0$ to the disk normal, in the X-Z plane, similar to simulation D of Mukherjee et al. (2018). A list of the relevant parameters of the simulations are presented in Table 8. The simulations have been performed with version 4.4 of the PLUTO hydrodynamic code (Mignone et al. 2007). We refer the reader to Mukherjee et al. (2016) and Mukherjee et al. (2018) for further details regarding the set-up of the simulations and the numerical schemes.

The jets strongly interact with the intervening clouds, generating turbulence and outflows, which are particularly strong in the vicinity of the jet head. This is shown in the top panel of Figure 12, where we present the column depth, mass weighted line of sight velocity of the dense gas (n > 1 cm$^{-3}$) and its mass weighted velocity dispersion at a time ~ 1 Myr since the launch of the jet. The images are viewed at an angle of $45^0$ with respect to the disk normal. The jet has partially cleared the central ~ 1 kpc of the dense disk, as seen from the column depth map. As the jet breaks out from the confines of the disk, it ablates and accelerates clouds, causing outflows. This can be seen in the map of the mass-weighted velocity field, in the form of mild distortions in the rotation profile along the jet, especially near its tip. This also results in enhanced velocity dispersion of the gas along the line of sight. The high dispersion of 400 km s$^{-1}$ near the jet tips is very similar to the observed results (e.g., see Murthy et al. 2022 and Figure 10 above).

Table 8. List of parameters for the hydrodynamic simulations

| Parameters | Value |
|---|---|
| Baryonic core radius ($r_B$) | 4 kpc |
| Baryonic velocity dispersion ($\sigma_B$) | 200 km s$^{-1}$ |
| Ratio of DM to Baryonic core radius ($\lambda$) | 5 |
| Ratio of DM to Baryonic velocity dispersion ($\kappa$) | 1.5 |
| Halo Temperature $T_H$ | $10^7$ K |
| Halo density at r = 0 ($n_{h0}$) | 0.1 cm$^{-3}$ |
| Turbulent velocity dispersion of warm clouds ($\sigma_t$) | 120 km s$^{-1}$ |
| Mean density of warm clouds ($n_{w0}$) | 100 cm$^{-3}$ |
| Ratio of azimuthal velocity of disk to Keplerian velocity ($\varepsilon$) | 0.9 |
| Jet Lorentz factor ($\Gamma$) | 3 |
| Ratio of jet rest mass energy to enthalpy ($\chi$) | 61.9 |
| Jet radius | 30 pc |
| Size of simulation grid | 640 x 640 x 640 |
| Resolution of simulation | 6.25 pc |

As the jet stream is deflected by the clouds, the jet plasma percolates through the ISM in all directions. The jet-induced flows drive shocks into the clouds. The shocked collisionally ionized gas is expected to cool down by emitting radiation over a broad range of wavelength and emission lines (e.g., see Meenakshi et al. 2022), including X-rays. In order to distinguish the hot gas shocked by the jet from the colder gas, we present 3D volume rendered images of the gas density (for n > 1 cm$^{-3}$) in two color palettes in the lower panels of Figure 12. Colder gas with T < 5000 K is shown in blue, whereas hotter gas with T > 5000 K is presented in yellow-red. It is clearly seen from the lower middle panel with a more edge-on view that the swept-up shock-heated gas lies above the colder gas. When viewed at a more inclined angle in the lower left panel, the shocked gas seems to be distributed both along the jet, as well as in the lateral directions to the jet. We simulate the expected morphology of the emission in the soft X-ray band (0.1-2 keV) due to cooling from the shock heated plasma in the lower right panel of Figure 12. The X-ray emissivities are obtained from tabulated non-equilibrium cooling functions calculated with MAPPINGS V (Sutherland et al. 2018). It can be seen that the shocked gas is expected to have strong emission in regions perpendicular to the jet and near its tip, very similar to the observed distribution of the soft X-ray emission found in this study (Figure 6). Thus, the simulations of jet-ISM interactions are qualitatively in agreement with the observed results as far as the dense gas kinematics and the hot gas distribution are concerned, further strengthening the interpretation that the softer regions in X-ray spectral studies are likely of thermal origin.

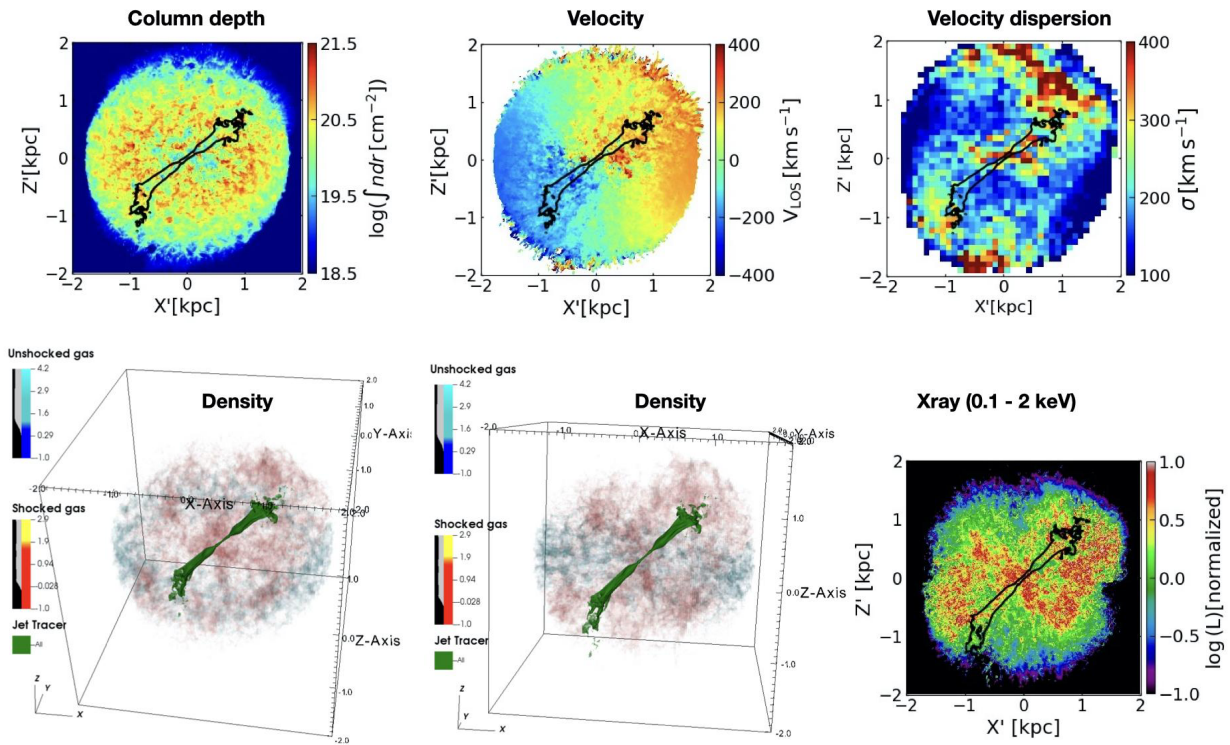

Figure 12. Gas distribution and kinematics as seen in simulations of a tilted (45°) $10^{44}$ erg s$^{-1}$ jet interacting with a gas-rich galactic disc. The jet lies in the *X-Z* plane. The 2D maps show projections along a line of sight at a 45° angle to the galactic discs (rotation around the *Y*-axis). Top left: Column density map. Top middle: Mass-weighted velocity dispersion. Top right: Mass-weighted velocity dispersion. Bottom left: A 3D volume rendering of the density of the simulated gas disk (with n > 1 cm$^{-3}$). Colder gas with T<5000 K, is shown in blue. Hotter gas with T > 5000 K, shown in red, represents dense gas shocked to high temperatures by the jet (in green). Bottom Middle: Same as in the left panel but at an orientation closer to an edge on view. Bottom right: A surface brightness map of the thermal X-ray emission in the energy range 0.1 - 2 keV from shocks in the dense clouds.

### 4.3 Non-thermal X-rays from the young radio source

The central 1 kpc scale radio source of B2 0258+35 with a jet power between $(0.82-1.3) \times 10^{44}$ erg $s^{-1}$ (Murthy et al. 2022) can contribute to the non-thermal X-ray and gamma-ray emission via inverse Compton scattering of the starlight and dust radiation off the relativistic electrons contained in the radio structures. Stawarz et al. (2008) modeled such contributions from the radio lobes at different evolutionary stages of a radio source varying total jet power and source size. Given their assumption about the optical and IR radiation fields, the model predicted X-ray luminosity for the radio source with a linear size of 1 kpc and jet power of $1 \times 10^{44}$ erg $s^{-1}$ (similar to the parameters of NGC1167) are below our NuStar upper limits. Our results are also consistent with the recent gamma-ray studies of young radio sources with Fermi (Principe et al. 2021) showing low detection rates.

### 4.4 X-rays – CO interactions

Recent studies of AGNs have shown that X-ray photons may have an effect on the CO emission of molecular clouds. In particular, CO 'cavities' have been reported in regions with intense X-ray emission (e.g., NGC 1068, Garcia Burillo et al. 2010; the Circinus galaxy, Kawamuro et al. 2019; ESO 428-G014, Feruglio et al. 2020; and NGC 2110, Rosario et al. 2019, Fabbiano et al. 2019). In ESO 428-G014, the CO-cavity is filled with warm molecular gas traced by $H_2$, directly suggesting an effect of the X-ray irradiation in the excitation of the CO lines. The X-ray radiation field may alter the chemistry of these regions, or fast X-ray producing shocks may warm up the CO molecules resulting in brighter emission in higher-J lines (see e.g., Feruglio et al 2020 and Kawamuro et al. 2020). The X-ray excitation of $H_2$ lines and suppression of CO emission has also been discussed in the context of NGC 4151 (Storchi-Bergmann et al 2009; Wang et al. 2011a).

We do not observe strong evidence of this phenomenon in NGC 1167. We have compared the Chandra full band and the HR images with the distribution of the cold molecular gas recently obtained from the NOEMA observations (Murthy et al. 2022). This comparison is shown in the left panel of Figure 10, where the X-ray iso-intensity contours are overlaid on the total intensity of CO(1-0). From the image it can be seen that the most intense region of the CO emission coincides with the termination of the southern jet (shown in the right panel of Figure 10). There is strong extended X-ray emission in this area, although the primary peak of X-ray emission coincides with the radio core/nucleus, which is to the NW of the CO peak. This could indicate perhaps an adverse effect of the X-ray photons on the CO emission in the nuclear region, but it is not a strong effect. The pronounced cavities observed in AGNs typically occur in the nuclear regions where the X-ray photon field is particularly intense. In NGC 1167, instead the nuclear source is relatively weak.

### 4.5 A 'turned-off' AGN?

Our combined Chandra and NuSTAR observations together strongly exclude the possibility of a highly obscured intrinsically luminous Compton Thick AGN. The faint detection of a nuclear source in the harder part of the Chandra energy band (see the hard localized emission in Figure 2), is consistent with a 0.3-7.0 keV luminosity of the nuclear source of $\sim 4 \times 10^{39}$ erg $s^{-1}$, based on the power-law fit of the Chandra spectrum (Table 7). However, this luminosity may be Compton scattered off an inner torus and so may represent only a small fraction of the total emitted nuclear luminosity. The $3 \sigma$ (8-24 keV) NuSTAR upper limit of $\sim <1 \times 10^{41}$ erg $s^{-1}$ (Section 3) gives a stronger constraint, as 8 - 24 keV photons can escape for $N_H$ values up to $\sim 10^{25}$ cm$^{-2}$. These $L_X$ values are in the low-luminosity AGN regime. Given the mass of the SMBH in NGC 1167 of 4.4 $\times$ $10^8$ M$_\odot$ (Kormendy & Ho 2013) and using the 8-24 keV limit and an AGN SED bolometric correction of 10 at $L/L_{Edd} \backsim < 0.01$ (Vasudevan & Fabian 2009; fig. 6 of Duras et al. 2020), corresponds to a bolometric luminosity of <1 x $10^{42}$ erg s$^{-1}$.

Is the extended X-ray luminosity of NGC 1167 solely due to jet-stimulated thermal emission? In this case, the X-ray emission would be entirely thermal, and NGC 1167 would be one of a kind source. In fact, previous Chandra results of nearby AGNs with disk-embedded jets show that photoionization is mostly responsible for the X-rays, although sizeable thermal components are present (see e.g., NGC 4151, Wang et al. 2011b; ESO 428-G014, Fabbiano et al. 2018a; IC 5063, Travascio et al. 2021). In ESO 428-G014, the non-nuclear extended X-ray luminosity due to photoionization is $\sim$5 x $10^{39}$ erg s$^{-1}$, while the thermal luminosity is $\sim$2 x $10^{39}$ erg s$^{-1}$. Similarly, in IC 5063 that has a total extended luminosity of $\sim 10^{40}$ erg s$^{-1}$, the thermal component is $\sim$5 x $10^{39}$ erg s$^{-1}$. If NGC 1167 is part of this cohort, it should also have a sizeable photoionized component of the extended X-ray emission. It would also be reasonable to compare its X-ray properties with those of other similar objects in the nearby universe studied with Chandra. In NGC 1167, the total X-ray luminosity of the extended emission detected with Chandra is $\sim$ 2 x $10^{40}$ erg s$^{-1}$ (0.3-7.0 keV). This is similar to the extended component luminosities of CT AGNs (see Fabbiano & Elvis 2022 for a summary), which however have $\sim$100 times or larger 'intrinsic' luminous nuclear sources in the 2-10 keV band (see Levenson et al. 2006). These CT AGNs have nuclear bolometric luminosities in the $10^{43}$-$10^{44}$ erg s$^{-1}$ range, way more than the <1 x $10^{42}$ erg s$^{-1}$ bolometric luminosity upper limit we measure for the nucleus of NGC 1167 from the newly obtained NuSTAR data.

We can speculate that NGC 1167 may be similar to the Hanny's Voorwerp AGN (IC 2497, Lintott et al. 2009, Sartori et al. 2018), an extended emission region remnant of past nuclear activity, but observed at a later evolutionary time, when the [OIII] emission has already faded. The recombination time of [OIII] is $\sim 10^4$ yrs (Lintott et al. 2009), much shorter than the lifespan of the extended X-ray emission ($\sim 10^7$ yr, see the discussion in Fabbiano & Elvis 2019 for the case of Hanny's Voorverp). Therefore, in NGC 1167 the optical signatures of high ionization may have faded away -consistent with its LINER spectrum (Gomes et al. 2016, Molina et al. 2018)- while the extended X-ray emission may still have the imprint of past AGN activity.

The multiwavelength observations of NGC 1167 / B2 0258+35 help constrain the lifecycle of the AGN. The total extent of the emission in the soft X-ray band is $\sim$ 1 kpc, well within the range observed in nearby luminous CT AGNs (Fabbiano & Elvis 2022). Given the light travel time, if this extended X-ray component is in part due to photoionization as in nearby AGNs with better S/N data, its radial extent suggests a past episode of sustained strong activity lasting $\sim$ 2 x $10^3$ yr. The relatively faint [OIII] emission suggests that the high accretion rate episode occurred earlier than $\sim 10^4$ yr ago, while the presence of 'young' radio jets, which may have switched on when the low accretion rate state occurred, puts the transition to the present low accretion rate regime between 0.4 Myr (Brienza et al. 2018) and 0.9 Myr (Giroletti et al. 2005) ago. These timescales are consistent with the accretion disk instability model of Czerny et al. (2009). Interestingly, a recent comparison of AGNs combining the SDSS and WISE MIR surveys finds an AGN population with luminous extended emission (from the [OIII] line as a tracer of the kpc-size narrow line region, SDSS) coupled with relatively faint nuclear dust emission (from the inner ~10pc, WISE), concluding that these AGNs have suffered a luminosity decline in the past $10^{3-4}$ years (Pflugradt et al. 2022).

4.6 What is the physical state of the nucleus of NGC 1167?

Given the mass of the SMBH in NGC 1167 of $4.4 \times 10^8$ M$\odot$ (Kormendy & Ho 2013), the measured nuclear luminosities (Section 4.5) suggest accretion rates ~$10^{-4}$ sub-Eddington (for a ~10% accretion efficiency). At these low accretion rates the AGN should currently be in a radiatively inefficient accretion flow (RIAF) state. In Galactic X-ray binaries, radio jets are produced in either the super-Eddington Very High state or, more commonly, in the low accretion rate (<0.01 $L_{Edd}$) RIAF state (Yuan and Narayan 2014). This is the 'hard spectral state' of Galactic black-hole X-ray binaries (Remillard and McClintock 2006). This may be the case in NGC 1167: beside hosting luminous young radio jets, NGC 1167 is exceptionally radio loud in its VLBA core. The ratio of the radio (5 GHz) VLBI core luminosity (Giroletti et al. 2005) to the Chandra nuclear (0.3 - 7 keV) luminosity is ~0.013, and compared to the NuSTAR (8-24 keV) upper limit is ~<0.33. These values compare with type 1 radio-loud AGN values of ~0.003 (Elvis et al. 1994.) The optical spectrum of the nuclear region also excludes the Seyfert activity detected in the extended X-ray emission regions of other CT AGNs (e.g., Maksym et al 2016; Ma et al. 2020), suggesting an entirely LINER source (Gomes et al. 2016). Sobolewska et al. (2011) model the spectral energy distribution of different types of AGNs and conclude that LINERs are in a similar 'hard spectral state' (although this may not universally apply to LINERs, Maoz 2007). Given the lack of current activity in the nucleus of NGC 1167 we suggest that a RIAF model may be compatible with the observations.

## 5. Summary and Conclusions

The deep 200 ks Chandra ACIS observation of NGC 1167 has revealed extended X-ray emission, mostly elongated in the direction of the B2 0258+35 radio jet, but also showing extent in the cross-jet direction. X-ray emission above that expected from a nuclear point source can be traced out to ~ 5" (~1.5 kpc) from the nucleus, encompassing the young radio jets.

In the 3-7 keV band there is a prominent X-ray peak in the 1/8 sub-pixel image, which we take as indication of nuclear emission and match astrometrically to the radio core. The centroid of the emission in the 0.3 – 3 keV image is displaced, suggesting nuclear obscuration.

Assuming a 70 Mpc distance, the total X-ray luminosity of the extended component is ~2 x $10^{40}$ erg s$^{-1}$, for a range of emission models (power-law + lines; thermal; photoionization). For the nuclear source we estimate an observed 0.3 - 7 keV luminosity of ~3 x $10^{39}$ erg s$^{-1}$.

The ACIS spectrum shows prominent soft emission with emission lines below 2 keV; spectral fits suggest NIX, Ne X, Mg XI and Si XIII, which have been found in the soft extended emission of other AGNs. While emission lines can be caused by either thermal or photoionized emission, the presence of Ne IX (0.915 keV) and Ne X (1.002 keV) lines suggests shock ionization, that may result from the interaction of a radio jet with the ISM (Wang et al. 2011b).

Given the limited statistics of the data we cannot distinguish between emission models, but the best fit parameters for both thermal and photoionization models are in the range of those found in CT AGNs observed with higher signal to noise (see Fabbiano & Elvis 2022 for a review). The kT values >1.5 keV implied by the thermal fits exceed the expectation for the hot gaseous components of E and S0 galaxies (Kim & Fabbiano 2015) and require energy input from an AGN.

At the higher energies the spectral count distribution is flat, reminiscent of the spectra of Compton Thick AGNs. There is no clear evidence of a prominent 6.4 keV Fe Kα line with EW~1000 eV typical of reflection spectra, although the statistical fit uncertainties cannot exclude one. The non-detection of NGC 1167 with our NuSTAR observations set a luminosity limit <1 x $10^{41}$ erg s$^{-1}$ (8-24 keV), excluding the presence of a typical CT AGN.

Both imaging and spectral analysis show an area of softer (lower energy) emission in the SE, partially coincident with the southern radio lobe. This soft region includes the area characterized by turbulence and fast outflow motions detected in the CO line emission (Murthy et al. 2022). If the X-ray emission is mainly thermal, the hot gas density in this soft region is ~ 2-3 times larger than in the surrounding harder emission region, and the cooling time ~10 times lower. The hot ISM may be compressed by the CO outflow, resulting in more efficient cooling. Also, the observed softer X-ray emission argues against intervening absorbing clouds, suggesting that the X-ray emitting gas resides between the CO clouds and the observer.

NGC1167 shows prominent cross-jet emission. Although it cannot be excluded that, at least in part, this emission may be due to a porous circumnuclear obscuring torus (see Fabbiano & Elvis 2022), lateral outflows have also been observed in the optical emission lines in at least one CT AGN (IC 5063, Maksym et al. 2021). The X-ray morphology of the extended X-ray emission of NGC 1167 could be consistent with the simulations of Mukherjee et al. (2018). We developed simulations tailored to NGC 1167, and found results qualitatively in agreement with the observations for both dense gas kinematics and hot gas distribution, strengthening the interpretation that the softer regions in X-ray spectral studies are likely of thermal origin.

NGC1167 is presently a LINER, but was an AGN in the past, given the properties of the extended X-ray emission and their similarity with those of CT AGN extended emission. We set a bolometric luminosity limit of <1 x $10^{42}$ erg s$^{-1}$ to the NGC 1167 AGN, consistent with accretion rates highly sub-Eddington (~$10^{-4}$ L/L$_{Edd}$). This low accretion rate would put the NGC 1167 AGN in the radiatively inefficient accretion flow (RIAF) state, which could also be responsible for the radio jets production (Yuan and Narayan, 2014).

The luminosity of the extended emission is comparable with that observed in $10^{43}$-$10^{44}$ erg s$^{-1}$ luminous CT AGNs, suggesting that we may be seeing the remnant of activity from an SMBH that was more active in the past, as reported in IC 2497 (the Hanny's Voorwerp AGN, Lintott et al. 2009). The total extent of the emission in the soft band is ~ 1 kpc, well within the range observed in nearby CT AGNs (Fabbiano & Elvis 2022). Given the light travel time, the radial extent suggests a past episode of sustained strong activity lasting ~ 2 x $10^3$ yr. This episode may have occurred more than ~$10^4$ yrs ago, given the constraint posed by the recombination time of [OIII]. The presence of 'young' radio jets would be consistent with this picture, since jets are expected to emerge for values of $L/L_{Edd}$ ~< 0.01 (Yuan and Narayan, 2014), consistent with our estimated $L/L_{Edd}$ value of ~$10^{-4}$.

**Acknowledgements**

We thank Francesca Civano for valuable comments on the manuscript. This work was partially supported by NASA guest observer grant GO1-22092X, and NASA contract NAS8-03060 (CXC). This research has made use of data obtained from the Chandra Data Archive and of the summary table (for NGC 1167) of the NASA/IPAC Extragalactic Database (NED). The NASA ADS bibliography service was used in this work. AYW is supported by JSPS KAKENHI Grant Number 19K03862. MB acknowledges support from the YCAA Prize Postdoctoral Fellowship. This work was partially performed at the Aspen Center for Physics, which is supported by National Science Foundation grant PHY-1607611.

**Software:** CIAO (Fruscione et al. 2006), Sherpa (Freeman et al. 2001; Doe et al. 2007; Burke et al. 2020), ChiPS (Germain et al. 2006), Cloudy (c08.01 Ferland et al. 1998), XSPEC (Arnaud 1996), NuSTARDAS (Perri et al. 2021), PLUTO (Mignone et al. 2007), MAPPINGS V (Sutherland et al. 2018).